\documentclass[nanomaterials,article,accept,moreauthors,pdftex]{Definitions/mdpi}

\usepackage{upgreek}
\firstpage{1}
\pubvolume{12}
\issuenum{19}
\articlenumber{3491}
\pubyear{2022}
\copyrightyear{2022}
\externaleditor{Academic Editor: M.V. Ramallo}
\datereceived{14 September 2022}
\dateaccepted{2 October 2022}
\datepublished{6 October 2022}
\hreflink{https://doi.org/10.3390/\linebreak{}nano12193491} % If needed use \linebreak
\pdfoutput=1

\Title{Ordered Bose Glass of Vortices in Superconducting YBa$_{2}$Cu$_{3}$O$_{7-\delta}$ Thin Films with a Periodic Pin Lattice Created by Focused Helium Ion Irradiation}

\TitleCitation{Ordered Bose Glass of Vortices in Superconducting YBa$_{2}$Cu$_{3}$O$_{7-\delta}$ Thin Films with a Periodic Pin Lattice Created by Focused Helium Ion Irradiation}

\Author{{Lucas Backmeister}
 $^{1}$, Bernd Aichner $^{1}$\href{https://orcid.org/0000-0002-9976-9877}{\orcidicon}, Max Karrer $^{2}$, Katja Wurster $^{2}$, Reinhold Kleiner $^{2}$\href{https://orcid.org/0000-0002-2819-9774}{\orcidicon}, Edward Goldobin $^{2}$, Dieter~Koelle $^{2}$\href{https://orcid.org/0000-0003-3948-2433}{\orcidicon}, Wolfgang Lang $^{1,}$*\href{https://orcid.org/0000-0001-8722-2674}{\orcidicon}}
\AuthorNames{Lucas Backmeister, Bernd Aichner, Max Karrer, Katja Wurster, Reinhold Kleiner, Edward Goldobin, Dieter Koelle, Wolfgang Lang}

\AuthorCitation{Backmeister, L.; Aichner, B.; Karrer, M.; Wurster, K: Kleiner, R.; Goldobin, E; Koelle, D.; Lang, W.}

\address{%
$^{1}$ \quad Faculty of Physics, University of Vienna, \mbox{A-1090 Wien, Austria}; {lucas.backmeister@univie.ac.at (L.B.); bernd.aichner@univie.ac.at (B.A.)}
\\
$^{2}$ \quad Physikalisches Institut, Center for Quantum Science (CQ) and LISA$^+$, Universit\"at T\"ubingen, \mbox{D-72076 T\"ubingen, Germany}; {max.karrer@uni-tuebingen.de (M.K.); katja.wurster@uni-tuebingen.de (K.W.); kleiner@uni-tuebingen.de (R.K.); gold@uni-tuebingen.de (E.G.); koelle@uni-tuebingen.de (D.K.)}
}

\corres{Correspondence: wolfgang.lang@univie.ac.at}
\abstract{
The defect-rich morphology of YBa$_{2}$Cu$_{3}$O$_{7-\delta}$ (YBCO) thin films leads to a glass-like arrangement of Abrikosov vortices which causes the resistance to disappear in vanishing current densities. This vortex glass consists of entangled vortex lines and is identified by a characteristic scaling of the voltage--current isotherms. Randomly distributed columnar defects stratify the vortex lines and lead to a Bose glass. Here, we report on the observation of an \emph{{ordered}
} Bose glass in a YBCO thin film with a hexagonal array of columnar defects with 30\,nm spacings. The periodic pinning landscape was engineered by a focused beam of 30\,keV He$^+$ ions  in a helium-ion microscope.
}

\keyword{copper-oxide superconductors; vortex glass; ordered Bose glass; vortex matching; voltage--current isotherms; helium-ion microscope}

\begin{document}

\section{Introduction}
%\textls[-15]{bla bla bla}

The copper oxide high-temperature superconductors (HTS) are in the extreme type-II limit, with a minor lower critical field $B_{c1}$ and a high upper critical field $B_{c2}$. The large difference in the critical fields spans a vast area in the phase diagram, the mixed state. The magnetic flux enters as Abrikosov vortices, quantized portions of flux $\Phi_0 = h/2e$, where $h$ is the Planck constant and $e$ is the elementary charge. The mixed state in HTS exhibits qualitatively new phenomenology \cite{BLAT94R} and is the predominant operating condition for most applications of these materials. Therefore, it is of utmost importance under which experimental conditions zero resistance, the hallmark of superconductivity, can be achieved. Moreover, the parameter space for utilizing superconductivity is limited by the critical temperature $T_c$, the upper critical field $B_{c2}$, and the critical current density $j_c$. The latter can be enhanced by various pinning mechanisms that block the dissipative motion of the vortices \cite{WORD17M,VALL22}. In HTS, the boundaries of this simple picture are substantially blurred by strong thermodynamic fluctuations of the superconducting order parameter.

In particular, the question of whether one can observe a genuine zero-resistance state at a finite temperature or only at zero temperature has raised much interest. Following the flux-creep theory of Anderson and Kim \cite{ANDE62,KIM62}, or the thermally-assisted flux-flow model (TAFF)~\cite{KES89}, one has to conclude that the resistance remains finite, even when the current density $j \rightarrow 0$.  Indeed, voltage--current ($V$-$I$) isotherms in the mixed state of many HTS reveal an ohmic behavior down to the lowest experimentally accessible voltages at temperatures not too far below $T_c$. This observation is attributed to TAFF. Numerical simulations reveal a rich variety of different dynamic phases when a vortex ensemble is driven over a background of correlated or random pinning defects \cite{REIC17R}.

The minuscule coherence lengths in HTS and the importance of intrinsic randomly distributed defects have triggered  theoretical proposals of a thermodynamic equilibrium phase with a glass-like arrangement of vortices. Different theories have been proposed for randomly distributed pinning defects, depending on their dimensionality: the vortex-glass (VG) model for point-like defects \cite{FISH89,FISH91} and the Bose-glass (BG) model for columnar defects penetrating the entire sample thickness \cite{NELS92,NELS93}.

The VG model has been confirmed by scaling of DC-current $V$-$I$ isotherms in a variety of HTS, among them YBa$_{2}$Cu$_{3}$O$_{7-\delta}$ (YBCO) in the shape of optimally doped thin films \cite{KOCH89}, oxygen-deficient films \cite{HOU94a}, ultrathin films \cite{DEKK92}, and single crystals \cite{GAMM91}. In addition, in the highly anisotropic compound Bi$_2$Sr$_2$CaCu$_2$O$_8$ (BSCCO-2212), an agreement with the VG theory was found in single crystals \cite{SAFA92a} and thin films \cite{YAMA94}. A transition from $D = 3$ to $D = 2$ VG scaling was reported in oxygen-depleted YBCO films \cite{SEFR99}. In addition, phase-resolved AC impedance measurements have provided another route to estimate the dynamic VG parameter $z_{VG}$ \cite{LANG96}.

The above considerations relate to defects in HTS that are not correlated along the crystallographic $c$ axis. In contrast, columnar defects (CDs) change the underlying physics, leading to a Bose glass \cite{NELS92,NELS93}. For instance, irradiation with swift heavy ions produces cylindrical channels of amorphous material that act as $c$-axis correlated pinning sites and evoke a BG behavior \cite{KRUS94a}. The fact that planar defects oriented parallel to the $c$ axis are ubiquitous in thin YBCO films can lead to inconsistent observations of VG \cite{WOLT93} or BG~\cite{SAFA96} behavior, depending on the details of the material's morphology. Moreover, earlier investigations of engineered nanoinclusions of different dimensionality have reported a crossover between VG and BG behavior \cite{HORI08}. For example, the latter was found for \emph{disordered} BaZrO$_3$ nanorods in YBCO.

The transitions between these various vortex phases, governed by temperature, magnetic field, and disorder, are long-standing issues. One aspect of the problem is that in earlier experiments, no periodic engineered pinning sites were available with spacings smaller than the London penetration depth and pronounced vortex pinning effects down to low temperatures.

Recent advances in nanopatterning of HTS by masked or focused light-ion irradiation~\cite{LANG06a,LANG09,PEDA10,AICH19} allow for the engineering of CDs with a periodic arrangement \cite{SWIE12,TRAS13,HAAG14,TRAS14,ZECH18a,AICH19,AICH20}. Using YBCO thin films with an unprecedented dense hexagonal lattice of CDs, created by focused He$^+$-ion-beam irradiation, allows us to observe a novel kind of glassy vortex correlations, which we will call an \emph{ordered Bose glass} (OBG). This paper explores the OBG phase  by measuring the $V$-$I$ isotherms at various temperatures and magnetic fields.

\section{Theoretical Background}
Both vortex and Bose glasses form below a magnetic-field-dependent glass temperature  $T_g(B)<T_c$ that marks a bifurcation between two essentially different domains of voltage--current ($V$-$I$) isotherms. While at $T>T_g(B)$ ohmic characteristics prevail down to vanishing $j$, the zero-resistance state emerges at $T<T_g(B)$ already at finite $j$. More importantly, the theories predict a critical scaling of several physical parameters at the continuous second-order phase transition between vortex or Bose glass and vortex liquid. In a VG, the central parameter is the glass correlation length $\xi_{VG} \propto |T-T_g|^{-\nu}$, which is determined by the size of glassy islands and of fluctuating vortex liquid droplets above and below $T_g(B)$. The lifetime of these fluctuations is $\tau \propto \xi_{VG}^z \propto |T-T_g|^{-\nu z}$.

The theoretical predictions can be experimentally verified by measuring $V$-$I$ isotherms near the VG transition and by comparing the data to the relation
\begin{equation} \label{eq:VG}
\frac{V}{I} \propto \xi_{VG}^{D-2-z} \mathfrak{F}_{\pm} \left(\frac{I \xi_{VG}^{D-1} \Phi_0}{k_B T} \right),
\end{equation}
where $\mathfrak{F}_{\pm}$ are two universal, yet unknown, characteristic functions of a specific VG system above ($\mathfrak{F}_+$) and below ($\mathfrak{F}_-$) $T_g(B)$, respectively, $D$ is the dimensionality of the vortex ensemble, and $k_B$ is the Boltzmann constant. By appropriately scaling a set of $V$-$I$ isotherms according to
\begin{equation} \label{eq:scalingVG}
(V/I) |1-T/T_g|^{\nu_{VG}(D-2-z_{VG})} = \mathfrak{F}_{\pm} [(I/T) |1-T/T_g|^{\nu_{VG}(1-D)}],
\end{equation}
collected at various temperatures and at fixed $B$, all curves collapse onto the two universal $\mathfrak{F}_{\pm}$ branches. The scaling is achieved by a proper choice of the parameters $T_g$, $\nu_{VG}$, and $z_{VG}$. The bifurcation line right at $T_g$ not only separates the branches $\mathfrak{F}_{\pm}$ but also obeys a power law.
\begin{equation} \label{eq:bifur}
(V/I)\mid_{T=T_g} \propto I^{(z_{VG}+2-D)/(D-1)}.
\end{equation}

Note that the BG theory predicts a similar scaling of the $V$-$I$ isotherms that can be cast into the same Equation~(\ref{eq:scalingVG}). Since a BG requires 3D correlations, the respective critical exponents of a 3D-VG can be transformed to those of a BG by \cite{WOLT93}
\begin{equation} \label{eq:coeff}
    \nu_{BG} = \frac{2\nu_{VG}}{3} \qquad \text{and} \qquad z_{BG} = \frac{3 z_{VG}+1}{2}.
\end{equation}

The VG and BG theories were developed for randomly arranged point and columnar defects, respectively. A new and unique situation arises when the CDs are periodically arranged. In an external magnetic field, applied parallel to the CDs, one or more magnetic flux quanta penetrate the CDs. The resulting magnetic commensurability (matching) fields~are
\begin{equation} \label{eq:match}
    B_m = m\frac{2\Phi_0}{\sqrt{3} a^2},
\end{equation}
where $m$ is a rational number, and $a$ denotes the lattice constant of a hexagonal CD lattice. At the matching fields $B_m$, when each CD can be filled on average by $m$ flux quanta, a significant change in the vortex dynamics can be expected. The most prominent effects are observed when $m$ is a natural number. At these matching fields, we observe the ordered Bose glass phase discussed below.

Tuning the magnetic field allows one to switch the vortex ensemble between VG and OBG. However, in the case of YBCO thin films with their strong intrinsic pinning by twin boundaries and screw dislocations, the VG, the BG, and the OBG are competing phases of increasingly frustrated disorder. At lower temperatures, the VG might evolve into a Bragg glass \cite{GIAM97}, where vortex dislocations are absent, and quasi-long-range translational order is preserved. Similarly, with vanishing disorder and at low temperatures, the OBG can transform into a vortex Mott insulator \cite{NELS93}, where the vortices condense in a commensurate arrangement with the CDs. Both Bragg glass and vortex Mott insulator differ from VG, BG, and OBG as they melt through a first-order transition.

%%%%%%%%%%%%%%%%%%%%%%%%%%%%%%%%%%%%%%%%%%
\section{Materials and Methods}
\subsection{Sample Preparation}
Thin YBCO films were epitaxially grown on (LaAlO$_3$)$_{0.3}$(Sr$_2$AlTaO$_6$)$_{0.7}$ (LSAT) substrates by pulsed laser deposition (PLD). The thickness of the film $t= (26.0 \pm 2.4)$\,nm used in this study was determined via Laue oscillations at the YBCO (001) Bragg peak. X-ray diffraction confirmed the excellent $c$-axis orientation of the film via the rocking curve of the YBCO (005) peak with a full width at half maximum (FWHM) of $0.08^\circ$.

Electrical contacts were established by first depositing a 20 nm-thick Au film employing in situ electron beam evaporation after the PLD process. Then, both the Au and the YBCO films were partially removed using Ar ion milling to form a bridge structure and the electrical contact pads. Afterward, parts of the Au layer were removed with Lugol's iodine to open a window for irradiating the bridge. The dimensions of the  YBCO microstrip are $8\,\upmu$m width and $40\,\upmu$m length with voltage probes separated by $20\,\upmu$m. The contact pads were connected by $50\,\upmu$m-thick Au wire and Ag paste to the cryostat's sample holder.

\subsection{Focused ion-beam irradiation}
The prepatterned YBCO microbridges were introduced into the Zeiss Orion NanoFab He-ion microscope (HIM) and aligned under low ion fluence. The focused He$^+$ ion beam is adjusted to an estimated average diameter of He$^+$ ion trajectories within the film of 9\,nm FWHM to avoid amorphization at the YBCO film's surface. An area of $36 \times 16\,\mu\text{m}^2$ was irradiated with a triangular spot lattice with distances $a = (30 \pm 0.6)$\,nm, using 30\,keV He$^+$ ions. The number of ions per spot is $10^4$, according to the dwell time of 3.2\,ms and a beam current of 0.5 pA. With this number of ions, $T_c$ can be locally suppressed, while only at higher ion fluence can an amorphization of the crystal structure be seen in high-resolution transmission electron microscopy {\cite{MULL19}}. More details about the formation of CDs by focused ion beam irradiation can be found elsewhere \cite{AICH19,AICH20}.

\subsection{Electrical measurements}
The electrical measurements were performed in a Physical Properties Measurement System (PPMS), equipped with a 9 T superconducting solenoid and a variable temperature insert (Quantum Design). The magnetic field was oriented perpendicular to the sample surface, and a Cernox resistor \cite{HEIN98} is used for in-field temperature control. The resistivity measurements were performed with a constant excitation current of $1\,\upmu$A in both polarities to exclude thermoelectric signals. The critical current is determined from isothermal $V$-$I$ measurements using a voltage criterion of 200\,nV. In addition, multiple $V$-$I$ curves were collected at fixed temperatures and stable magnetic fields, which were limited to $100\,\upmu$V to avoid heating effects. All data were collected by computer-controlled data acquisition.

%%%%%%%%%%%%%%%%%%%%%%%%%%%%%%%%%%%%%%%%%%
\section{Results and Discussion} \label{sec:results}

\begin{figure}[H]%[t]
\includegraphics*[width=\columnwidth]{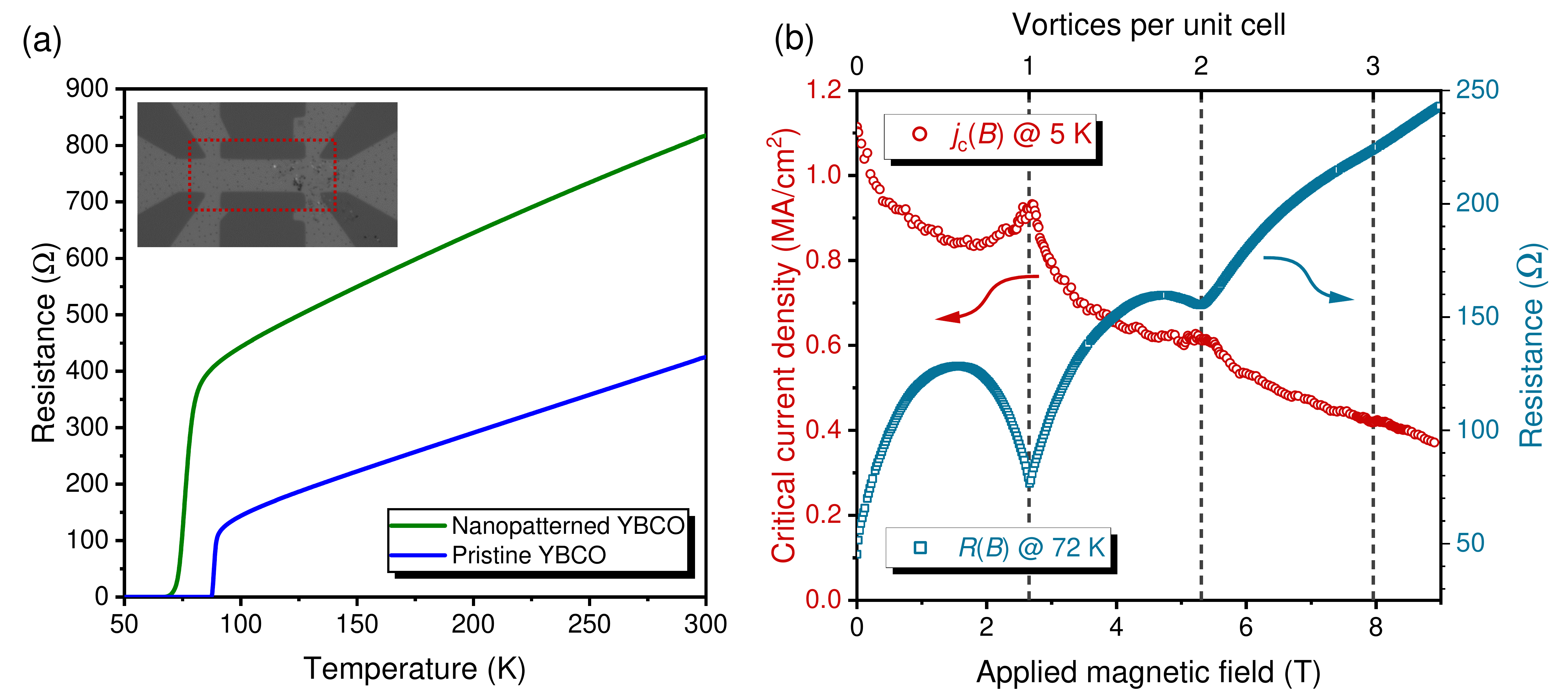}
\caption[]{(\textbf{a}) Resistances of a YBCO thin film with a hexagonal array of 30 nm spaced columnar defects and an unirradiated reference bridge fabricated on the same substrate. Inset: Optical microscopy picture of the bridge taken after irradiation. The red dotted rectangle marks the irradiated area of $36 \times 16\,\upmu\text{m}^2$. (\textbf{b}) Critical current density at 5\,K and resistance at 72\,K of the irradiated bridge. The broken lines indicate the number of vortices per unit cell of the hexagonal CD lattice, corresponding to the matching fields $B_m = m \times 2.653$\,T, which were calculated from Equation~(\ref{eq:match}) using the nominal geometry of the irradiation pattern.}
\label{fig:transport}
\end{figure}

The temperature dependence of the resistance $R$ of a YBCO microbridge after patterning with a hexagonal array of 30\,nm spaced CDs is shown in Figure~\ref{fig:transport}a, together with the $R(T)$ curve for an unirradiated microbridge, as a reference sample. The inset displays an optical microscopy picture of the YBCO bridge \textls[-5]{taken after irradiation in the HIM. The red dotted rectangle marks the irradiated area, extending over the entire bridge between the voltage probes and containing about $7.4 \times 10^5$ CDs according to the irradiation protocol. Even at much higher magnification in the HIM, we cannot identify the signatures of the CDs. Actually, ion fluence was deliberately chosen such that mainly oxygen atoms are displaced, but amorphization of the material is avoided. It is only at more than an order of magnitude higher fluence that we can visualize local destruction of the crystal structure \cite{AICH19,MULL19}.}

The non-irradiated YBCO bridge, prepared as a reference on the same substrate, has a $T_c = 88.4$\,K (defined as the inflection point). The $T_c$ in very thin films is generally slightly lower than in single crystals due to the strain imposed by the substrate. Focused ion beam irradiation in the HIM causes a reduction to $T_c = 76.0$\,K. The $T_c$ suppression is much less than previously reported for masked He$^+$ ion irradiation of CDs with larger diameters and spacings \cite{SWIE12,HAAG14}. We suspect that the scattering of individual ions away from the CDs creates a small number of point defects between the CDs, which are known to reduce $T_c$. The similar slopes of the pristine and nanopatterned bridges indicate that irradiation does not affect the charge carrier density in the inter-CD regions. On the other hand, the offset of a linear extrapolation of the normal state resistance at zero temperature is much higher in the irradiated sample, which is attributed to enhanced defect scattering \cite{SEFR01,LANG10R}.

Figure~\ref{fig:transport}b presents the pronounced vortex commensurability effects in $j_c$ at 5\,K and $R$ at 72\,K of the nanopatterned YBCO bridge as a function of the magnetic field applied orthogonal to the sample surface. The critical current density serves as a static probe, and a well-developed peak is centered around the matching field $B_1 = 2.653$\,T. It perfectly agrees with $B_1$ calculated from Equation~(\ref{eq:match}) inserting the $a = 30$\,nm of the irradiation protocol in the HIM. The noticeable dip in the flux-flow resistance indicates that commensurability is not lost in approaching the vortex liquid regime and might indicate a plastic-flow phase~\cite{REIC97a}. It was established previously by the angular dependence of $B_m$ in tilted magnetic fields~\cite{TRAS13,AICH20} that the CDs indeed act as one-dimensional pinning lines in such systems. In addition, a less prominent peak of the critical current density and a corresponding dip in the resistance is present at $B_2 = 5.31$\,T, i.e., with an arrangement of two vortices per unit cell of the CD lattice. These observations justify studying $V$-$I$ curves further to explore the critical scaling mentioned earlier.

\begin{figure}[H]%[t]
\includegraphics*[width=\columnwidth]{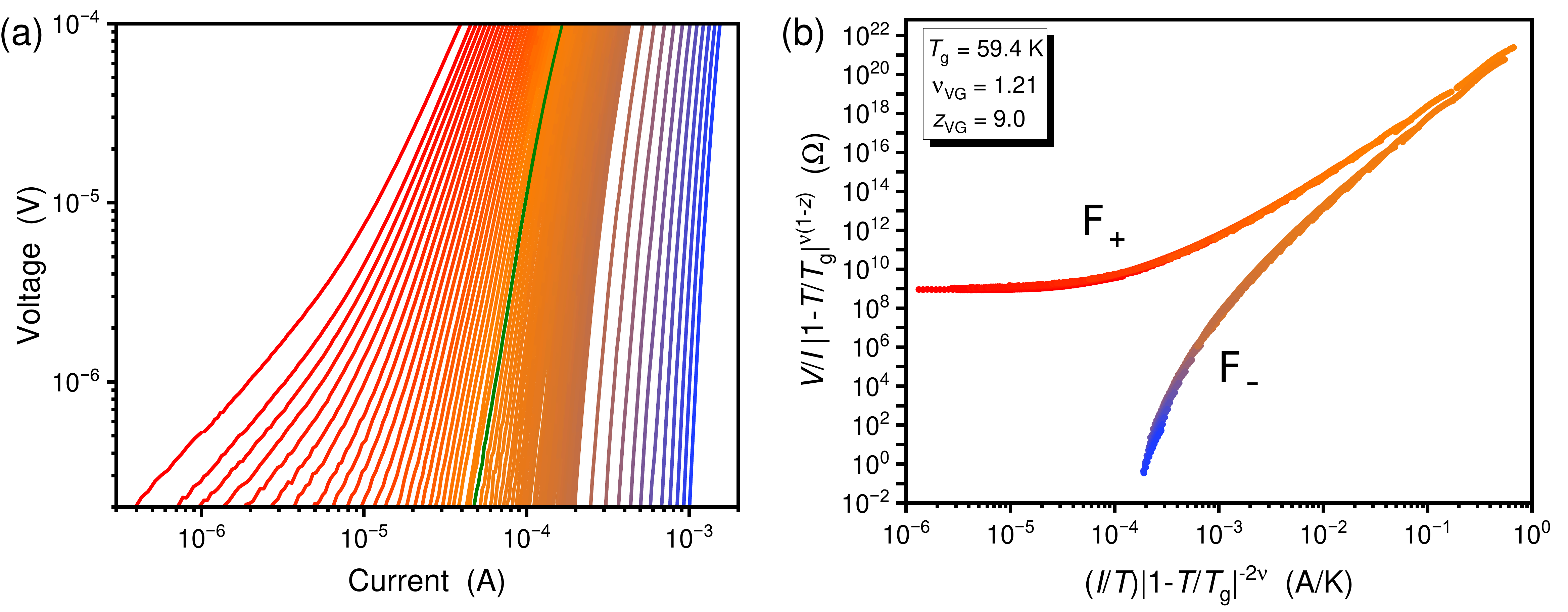}
\caption[]{(\textbf{a}) $V$-$I$ isotherms at 66\,K (red), 65.5 to 52\,K in 0.25\,K steps, and 50  to 30\,K (blue) in 2\,K steps of the nanopatterned sample at the matching field $B_1 = 2.653$\,T. The green line indicates the isotherm at 59.5\,K, which is closest to the glass temperature $T_g = 59.4$\,K.  (\textbf{b}) $V/I$ vs. $I$ isotherms plotted according to the VG scaling of Equation~(\ref{eq:scalingVG}). The isotherms collapse onto the two universal functions $\mathfrak{F}_+$ above and $\mathfrak{F}_-$ below $T_g$, respectively. The colors of the data points represent the temperature and are the same as in panel (\textbf{a}).}
\label{fig:IV}
\end{figure}

The $V$-$I$ characteristics were recorded in static magnetic fields ranging from 1.0 to 8.0\,T and with temperatures in the vicinity of the glass temperature as the variable parameter. For example, the isotherms for the matching field $B_1 = 2.653$\,T are presented in Figure~\ref{fig:IV}a. At 66\,K (red lines) and above (not shown), an ohmic behavior is observed in the low-current limit and attributed to TAFF. At lower temperatures, non-linearity emerges, and a power-law behavior can be observed over the entire current range. The $V$-$I$ characteristics at $T < T_g$ (to the right of the green line) have negative curvatures that hint at a vanishing resistance for a particular critical current. All these observations match the predictions of both the VG and BG theories and Equation~(\ref{eq:VG}).

Following the scaling theories, Figure~\ref{fig:IV}b shows the collapse of the $V/I$ vs. $I$ data deduced from panel (a) according to Equation~(\ref{eq:scalingVG}). Two branches, corresponding to the universal functions $\mathfrak{F}_+$ and $\mathfrak{F}_-$ are built up from the various isotherms. The results appear qualitatively similar to previous works, except for the dynamic parameter $z_{VG} = 9.0$ that falls outside the reasonable range predicted by the VG theory and also differs from previous experimental findings.

\begin{figure}[H]%[t]
\includegraphics*[width=\columnwidth]{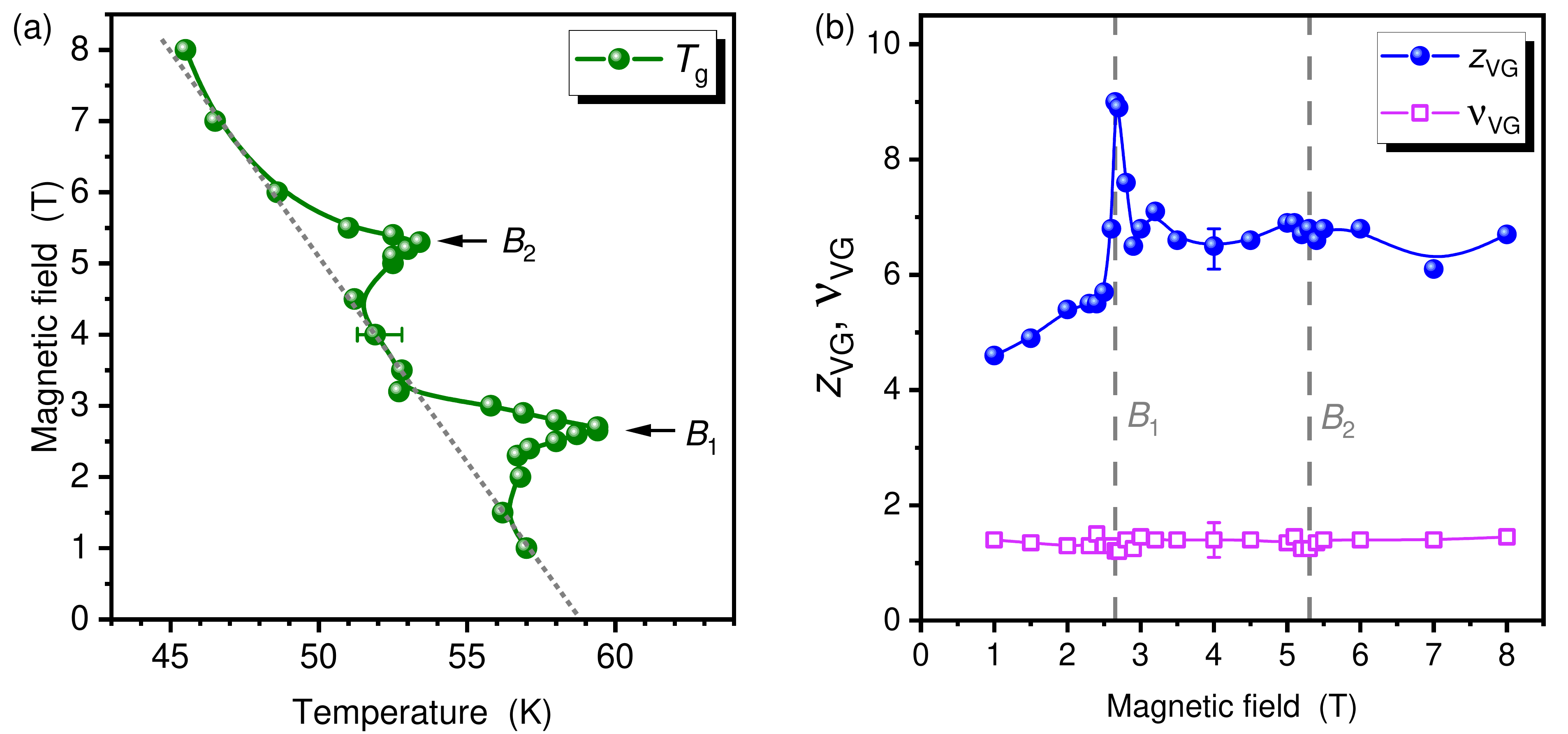}
\caption[]{Scaling parameters of the nanopatterned YBCO film. The representative error bars indicate the uncertainty caused by interdependence between the fit parameters $T_g$, $z_{VG}$, and $\nu_{VG}$. The size of the symbols represents the uncertainty of the parameter when the other two parameters are kept fixed. Solid lines are guides to the eye. (\textbf{a}) The phase boundary between the vortex liquid and the glass phase is represented by the glass temperature $T_g$ as a function of the external magnetic field. The gray dotted line symbolizes a linear trend of $T_g$. (\textbf{b}) Parameters from the scaling analyses according to Equation~(\ref{eq:scalingVG}).}
\label{fig:params}
\end{figure}

Figure~\ref{fig:params}a illustrates the systematic change of the glass temperature $T_g$ with the applied magnetic field. Pristine YBCO shows an almost linear decrease of $T_g$ with the magnetic field~\cite{KOCH89}. This trend, indicated by the dotted line, can also be observed in the nanopatterned sample, as long as the applied magnetic field is not commensurable. However, at $B_1$ and, to a lesser extent, $B_2$, peaks of $T_g$ are visible.
The scaling parameters $\nu_{VG}$ and $z_{VG}$ in different magnetic fields behave very diverse, as demonstrated in Figure~\ref{fig:params}b. On the one hand, the parameter $\nu_{VG}$ is independent of the magnetic field and has the typical value $\nu_{VG} = 1.3 \pm 0.2$, which is also reported for pristine YBCO~\cite{LANG96}. On the other hand, $z_{VG} = 6.5 \pm 0.5$ at $3\,\text{T} < B < 8\,\text{T}$ but exhibits a narrow peak at $B_1$ with a maximum $z_{VG} = 9.0$. This value of $z_{VG}$ lies clearly outside the 3-dimensional (3D) VG scaling range. Interestingly, $z_{VG}$ does not peak at $B_2$, and the minor broad hump is within error limits. We can imagine two possible reasons: either the pinning force of the second vortex in a CD is rather small so that it can easily escape, or the second vortex is already magnetically caged at an interstitial position. In both scenarios, the dynamics of moving vortices will be dominated by the disordered defects that are present between the CDs and result in a similar $z_{VG}$ as in off-matching fields.

In Figure~\ref{fig:params}, we have introduced two different error estimates. The size of the symbols indicates the respective uncertainty of $T_g$, $\nu_{VG}$, and $z_{VG}$ when the other two parameters are kept fixed. There is, however, interdependence between the choice of parameters, mainly between $T_g$ and $z_{VG}$. An increase of $T_g$ relates to a decrease of $z_{VG}$, and vice versa. The range in which reasonable scaling collapses can be achieved is marked by error bars. It is crucial that the significant matching effect at $B_1$ is represented by an increase of \emph{both} $T_g$ and $z_{VG}$. Hence, despite any uncertainties with the collapsing $V/I$ vs. $I$ curves, a marked change in the vortex dynamics is evident.

When converting the critical exponents to those of the BG theory by Equation~(\ref{eq:coeff}), we find $\nu_{BG} = 0.81$ and $z_{BG} = 14$ at $B_1$. Note that both theories predict comparable values for $\nu$, and, as discussed above, we do not observe any features of $\nu_{VG}$ at the matching fields in the nanopatterned sample. Conversely, the value for $z_{BG} = 14 $ is significantly above previous results ($z_{BG}$$\sim$9) in YBCO with incommensurate BaZrO$_3$ (BZO) nanorods \cite{HORI08}. We attribute this enhancement of $z_{BG}$ to the periodic arrangement of the CDs in our samples and thus to the OBG phase. A theoretical analysis reported \cite{GIAM97} that in the quasi-long-range ordered Bragg glass, the $V$-$I$ isotherms are steeper than in the VG. According to Equation~({\ref{eq:bifur}}), steeper isotherms are connected with a larger $z$, which also points to an increase of $z$ by the frustration of disorder and is in accordance with our observations.

\begin{figure}[H]%[t]
\includegraphics*[width=0.6\columnwidth]{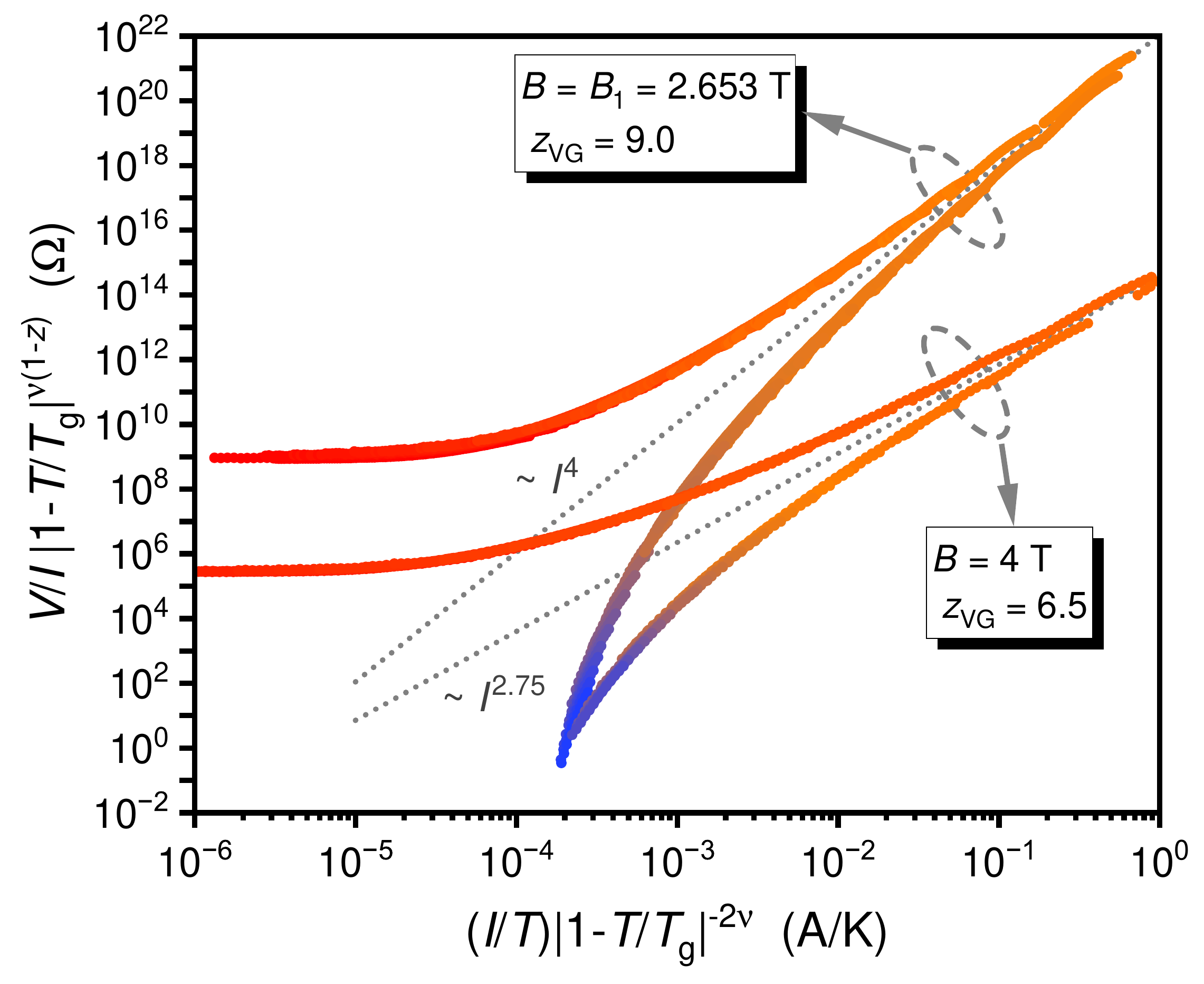}
\caption[]{Comparison of the collapsed $V/I$ vs. $I$ curves at the first matching field $B_1 = 2.653$\,T and at an off-matching field $B = 4$\,T. The dotted lines indicate the bifurcation and have different exponents. Color coding is the same as in Figure~\ref{fig:IV}.}
\label{fig:compare}
\end{figure}

In Figure~\ref{fig:compare}, the collapsed curves at the commensurability field $B_1$ and an off-matching field of 4\,T are compared. The dotted lines indicate the bifurcation between the universal functions $\mathfrak{F}_{\pm}$. Remarkably, these bifurcation lines have significantly different slopes, which are, according to Equation~(\ref{eq:bifur}), determined solely by the dynamic scaling parameter $z_{VG}$ (or $z_{BG}$). Thus, steeper $V/I$ vs. $I$ isotherms at $B_1$ and, at the same time, $z_{BG}$ above the typical values for a disordered Bose glass strongly support our proposal of an \emph{ordered} Bose glass phase.

One might raise concerns about a `true' 3D nature of vortices in very thin films rendering 2D scaling theories more appropriate \cite{DEKK92}. Taking the distance of vortices $l \simeq \sqrt{\Phi_0/B}$, a 3D vortex glass is expected when $l \lesssim t$. For our film with $t = 26$\,nm, the condition for 3D is fulfilled for $B \gtrsim B_1$.

As a cross-check, we can evaluate the critical parameters for 2D scaling.
It follows from Equation~(\ref{eq:scalingVG}) that $\nu_{VG}^{2D} = 2 \nu_{VG}^{3D}$ and $z_{VG}^{2D} = (z_{VG}^{3D}-1)/2$. Thus, the same collapse of the $V$-$I$ isotherms can be achieved with the adapted parameters of the 2D model. However, $\nu_{VG}^{2D} = 2.6 \pm 0.4$ would be outside the expected range between 1 and 2, and, moreover, $z_{VG}^{2D} = 2.75 \pm 0.25$ at $3\,\text{T} < B < 8\,\text{T}$ is too low for a VG. Only at $B_1$, $z_{VG}^{2D} = 4.0$ would be an acceptable value. A typical 2D vortex ensemble is weakly coupled pancake vortices. Assuming that such a 2D situation arises by introducing columnar pinning defects would be counterintuitive. We thus conclude that a 2D glass scaling theory is incompatible with our observations.

Our results can be compared with previous findings. In pristine YBCO thin films at a magnetic field $B = 2$\,T, 3D scalings with $\nu_{VG} = 1.7, z_{VG} = 4.9$ \cite{KOCH89} and $\nu_{VG} = 1.3 \pm 0.1, z_{VG} = 4.8 \pm 0.3$ \cite{LANG96}, respectively, have been reported. These values correspond well to our present results for $B < B_1$. Theoretically, the two critical parameters are expected to be independent of the applied magnetic field. However, at $B > B_1$, we observe an increase to $z_{VG} \simeq 6.5$, which is slightly above the typical bandwidth for VG scaling.

A possible reason is that a fraction $B_1/B$ of vortices are bound in the CDs, whereas additional flux penetrates as interstitial vortices. Such a landscape of correlated quenched disorder by the CDs and various disordered pinning sites (twin boundaries and screw dislocations) between them could lead to more complex vortex dynamics and thus to $z_{VG}$ values between those at low fields and the high one observed at $B_1$.

Similar considerations lead to the difference between a vortex Mott insulator and the present situation of an OBG. A vortex Mott insulator should show no glassy behavior since vortex lines can equidistantly hop from one CD to the next \cite{SORE17}. In an OBG, however, the disordered intrinsic defects between the CDs can also trap vortices and induce glassiness.

In YBCO thin films that were large-area irradiated (i.e., without lateral modulation of the fluence) with 80\,keV He$^+$ ions, $\nu_{VG} = 1.4, z_{VG} = 4.75$ have been reported, and no difference was found between irradiated and pristine films \cite{SEFR01}. Hence, we conclude that it must be the defect pattern created by focused He$^+$ ion beam irradiation in our sample that leads to the present observations.

Remarkably, a study of thin YBCO films after irradiation with 100\,keV O$^+$ ions through a masking layer penetrated by a square lattice of holes with 120\,nm spacings found 2D scaling of the $V$-$I$ isotherms, indicating reduced vortex-glass correlations along the vortex line \cite{TRAS13}. In addition, no significant change of the critical exponents $\nu_{VG}$ and $z_{VG}$ has been found at the matching field, yet there is a slight enhancement of the vortex glass temperature. Although there are several differences in our study, such as hexagonal versus square CD patterns and the different nature and diameters of the CDs, we believe that the main reason is the CD's density. The condition $l \lesssim t$ discussed above is not met, which might indeed lead to a 2D-VG behavior.

%%%%%%%%%%%%%%%%%%%%%%%%%%%%%%%%%%%%%%%%%%
\section{Conclusions}

Thin films of the copper-oxide superconductor YBCO contain many intrinsic defects, such as point defects, twinning phase boundaries, and screw dislocations, that act as randomly distributed pinning sites. An ultradense hexagonal arrangement of columnar defects created by focused He$^+$ ion irradiation can overcome the uncorrelated pinning in YBCO thin films. We have found that in such a sample, the vortex dynamics and the relative strength of periodic and uncorrelated pinning sites can be tuned by the applied magnetic field. Furthermore, critical scaling relations of the voltage--current isotherms in constant magnetic fields yield parameters that distinguish between vortex glass and an ordered Bose glass behavior. The latter can emerge from a vortex Mott insulator when thermal energy and disorder weaken the vortex correlations, as is the case in our experiments. Accordingly, we observe the ordered Bose glass phase exactly at the matching field, where every columnar defect can be filled by one flux quantum.

Still, many intriguing issues exist in the vortex physics of copper-oxide superconductors. The fabrication of engineered pinning landscapes in copper-oxide superconductors with controlled geometry and tens of nm spacings may raise the investigations of vortex systems and their various phases originating from the competition of pinning and elastic vortex forces, disorder, and fluctuations to a higher level.

\vspace{6pt}

%%%%%%%%%%%%%%%%%%%%%%%%%%%%%%%%%%%%%%%%%%
\authorcontributions{W.L., D.K., R.K., and E.G. conceived and supervised the experiments, K.W. grew the film, M.K. patterned the film and performed the focused ion beam irradiation, L.B., B.A., and W.L. performed the transport measurements, L.B. and W.L. analyzed the data, all authors discussed the results and contributed to writing the paper. All authors have read and agreed to the published version of the manuscript.}

\funding{{Open Access Funding by the Austrian Science Fund (FWF)}.
%YES, but to be grammatically correct, I added a few words below.
The research was funded by a joint project of the Austrian Science Fund (FWF), grant I4865-N, and the German Research Foundation (DFG), grant KO~1303/16-1, and it is based upon work from COST Actions CA16218 (NANOCOHYBRI), CA19108 (Hi-SCALE), and CA19140 (FIT4NANO), supported by COST (European Cooperation in Science and Technology).}

\dataavailability{The data presented in this study are available on reasonable request from the corresponding author.}

%\acknowledgments{In this section you can acknowledge any support given which is not covered by the author contribution or funding sections. This may include administrative and technical support, or donations in kind (e.g., materials used for experiments).}

\conflictsofinterest{The authors declare no conflict of interest.}

%%begin novalidate % to avoid overleaf error report with paracol package
%\end{paracol}
%%end novalidate

\begin{adjustwidth}{-\extralength}{0cm}
%\centering %% If there is a figure in wide page, please release command \centering
\reftitle{References}

%\externalbibliography{yes}
%\bibliography{OBG}
\end{adjustwidth}

\end{document}